%% file: main.tex
\documentclass[11pt]{article}
\usepackage{times}
\usepackage{geometry}
\usepackage[utf8]{inputenc}
\usepackage{enumitem,amssymb}
\usepackage{ragged2e}
\newlist{thematic}{itemize}{8}
\setlist[thematic]{label=$\square$}
\usepackage{pifont}
\usepackage{fancyhdr}
\usepackage[USenglish]{babel}
\usepackage[utf8]{inputenc}
\usepackage[T1]{fontenc}
\usepackage{epsfig}
\usepackage{wrapfig}
\usepackage{color}
\usepackage{comment}
\usepackage{MnSymbol,wasysym}
\usepackage{framed}

\usepackage[vflt]{floatflt}
\usepackage{longtable}
\usepackage{caption}
\usepackage{jheppub}   
\usepackage[dvipsnames,svgnames,x11names]{xcolor}
\usepackage[all]{hypcap} 
\usepackage{amssymb,amsmath}
\usepackage{longtable}
\usepackage{amssymb}
\usepackage{amsmath, xspace}
\usepackage{graphics,graphicx} 
\usepackage{rotating}
\usepackage{color}
\usepackage{xcolor}
\usepackage{multirow}
\usepackage{subfigure}
\usepackage{sectsty}
\usepackage[outercaption]{sidecap}
\sidecaptionvpos{figure}{c}

\input{macros}

\usepackage[backend=biber,style=authoryear,citestyle=numeric-comp,bibstyle=science,hyperref=true,uniquename=mininit,uniquelist=false,natbib=false,defernumbers=true,eprint=false,sorting=none,maxcitenames=1,maxbibnames=1,]{biblatex}
\makeatletter
\newcommand*{\blx@fnpct@movefor}{}
\newcommand*{\DeclareFootnoteMovePunct}{%
  \@ifstar
    {\let\blx@fnpct@movefor\@empty}
    {}
  \blx@def@fnpct@movefor}
\newcommand*{\blx@def@fnpct@movefor}{%
  \def\do##1{\blx@thecheckpunct{\listadd{\blx@fnpct@movefor}}##1}%
  \docsvlist}
\DeclareFootnoteMovePunct{.,{,}}
\newlength{\blx@fnpct@movelength}
\newcommand*{\blx@fnpct@footnotemover}[1]{%
  #1%
  \ifinlist{#1}{\blx@fnpct@movefor}
    {\settowidth{\blx@fnpct@movelength}{#1}%
     \hspace{-1.\blx@fnpct@movelength}}
    {}%
}
\protected\csedef{blx@acitei@superscript}#1#2#3#4#5{%
  \begingroup
  \blx@citeinit
    \noexpand\blx@fnpct@footnotemover{#5}%
    \blxcitecmd{supercite#1}{#2}{#3}{#4}{}%
  \endgroup}%
\protected\csedef{blx@macitei@superscript}#1#2#3{%
  \noexpand\blx@fnpct@footnotemover{#3}%
  #1{#2}%
  \endgroup}
\makeatother

\DeclareFieldFormat[article]{volume}{#1}
\DeclareFieldFormat[article]{journaltitle}{#1}

\AtEveryBibitem{\clearfield{doi}}
\AtEveryCitekey{\clearfield{doi}}

\AtEveryBibitem{\clearfield{url}}
\AtEveryCitekey{\clearfield{url}}

\AtEveryBibitem{\clearfield{month}}
\AtEveryCitekey{\clearfield{month}}

\DeclareBibliographyDriver{arxiv}{%
    \printnames{author}%
    \newunit\newblock%
    \printfield{eprint}%
    \setunit{\space}%
    \printfield{year}
}

\DeclareFieldFormat[arxiv]{title}{{#1}}
\DeclareFieldFormat[arxiv]{year}{({#1})}
\DeclareFieldFormat[arxiv]{eprint}{arXiv:{#1}}

\renewbibmacro*{name:andothers}{%
  \ifboolexpr
    {
      test {\ifnumequal{\value{listcount}}{\value{liststop}}}
      and
      test \ifmorenames
    }
    {
      \ifnumgreater{\value{liststop}}{1}
        {\finalandcomma}
        {}%
      \andothersdelim
      \bibstring{andothers}%
    }
    {}%
}

\DeclareFieldFormat{labelnumberwidth}{\textbf{[#1]}}

\defbibenvironment{bibliography}
  {}
  {}
  {\addspace
   \printtext[labelnumberwidth]{%
     \printfield{prefixnumber}%
     \printfield{labelnumber}}%
   \addhighpenspace}
   
\usepackage[most]{tcolorbox}

\begin{document}
\thispagestyle{empty}
\pagenumbering{gobble}

\Huge
\noindent Mass Assembly and Chemical Complexity in the Milky Way\\
\bigskip
\normalsize

\noindent\textbf{Authors:} \\
Pamela Klaassen (UK Astronomy Technology Centre, UK);\\
David Eden (University of Bath, UK); \\
Alessio Traficante (INAF-IAPS, Italy); \\ 

\noindent\textbf{Alphabetical Co-Authors}: \textit{Henrik Beuther} (Max Planck Institute for Astronomy, Germany), 
\textit{Maite Beltr\'{a}n} (INAF–Osservatorio Astrofisico di Arcetri, Italy),
\textit{Caroline Bot} (Observatoire Astronomique de Strasbourg, FR),
\textit{Elias Brinks} (University of Hertfordshire, UK),
\textit{Laura Colzi} (Centro de Astrobiología, CSIC–INTA, Spain),
\textit{Timea Csengeri} (Max Planck Institute for Radio Astronomy, Germany),
\textit{Antoine Gusdorf} (Laboratoire de Physique de l’ENS, France),
\textit{Doug Johnstone} (NRC Herzberg Astronomy \& Astrophysics, Canada),
\textit{Jes K. J{\o}rgensen} (Niels Bohr Institute, University of Copenhagen, Denmark),
\textit{Jonathan Marshall} (Academia Sinica Institute of Astronomy \& Astrophysics, Taiwan),
\textit{Elena Redaelli} (European Southern Observatory, Germany),
\textit{Víctor M. Rivilla} (Centro de Astrobiología, CSIC–INTA, Spain),
\textit{Thomas Stanke} (Max Planck Institute for Extraterrestrial Physics, Germany)

\vspace{12pt}
\noindent \textbf{Science Keywords:}
Interstellar Medium: Chemistry, Clouds, Star Formation, Kinematics and Dynamics\\

 \captionsetup{labelformat=empty}
\begin{figure}[h]
   \centering
\includegraphics[width=.6\textwidth]{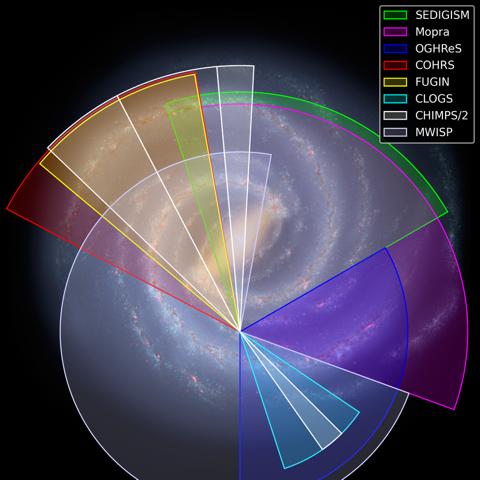} 
\rotatebox{90}{\includegraphics[width=.600\textwidth]{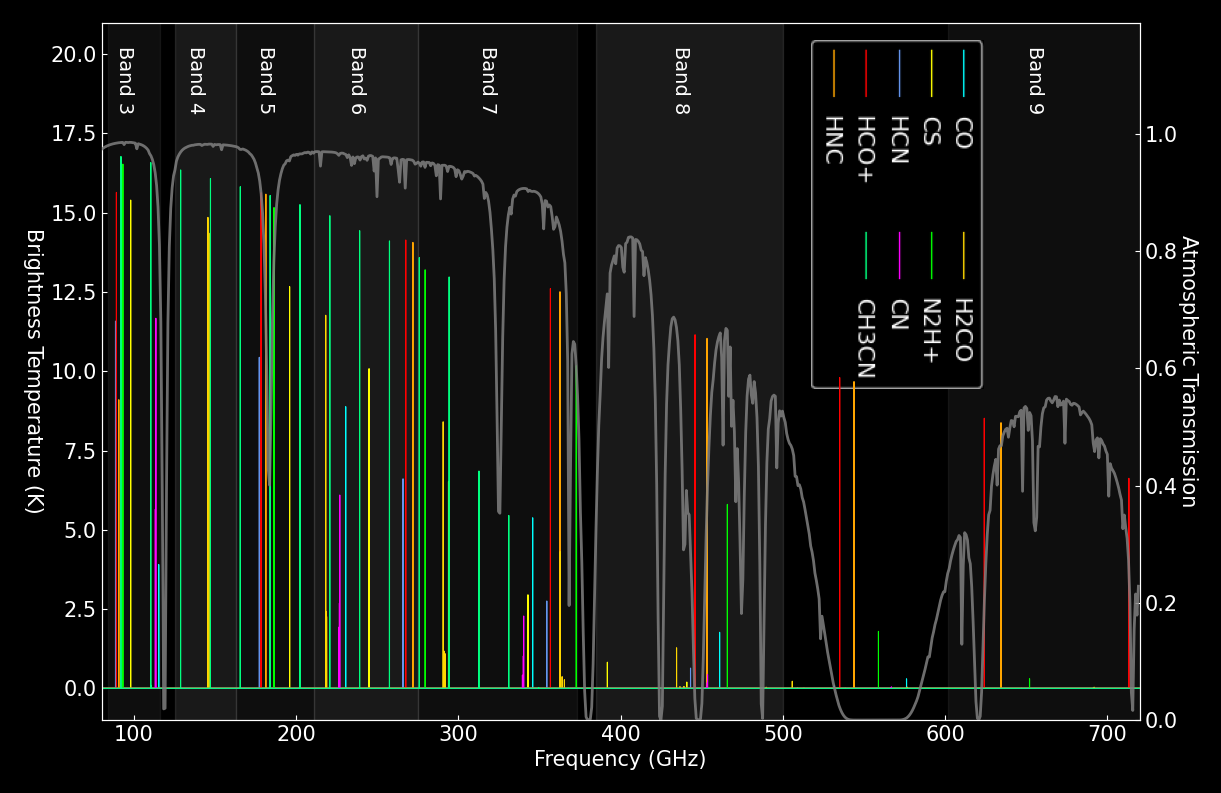}}
   \caption{\textit{Left:} Current state of Galactic Plane Line surveys of a single species (CO). View modelled after Figure 1 of \cite{Schuller21} (Background Image Credit: R. Hurt [NASA/JPL/Caltech]). \textit{Right:} What spectral lines could be detected homogeneously across the plane with large scale studies with next generation facilities.  }
   \label{fig:overview}
   
\end{figure}
\vspace{-15mm}

\pagebreak
\setcounter{page}{1}
\pagenumbering{arabic}

\begin{tcolorbox}[colback=blue!10!white]
(Sub-)millimeter spectral lines can be used not only to understand the chemical complexity and enrichment history of an observed portion of our Galaxy, but with spectrally resolved lines, they reveal the physical conditions, dynamics, and even the ionisation state and magnetic field strengths of the gas component of our Galaxy. They are prime tracers of mass assembly and structure formation across scales.
\end{tcolorbox}


\vspace{-18pt}
\section*{Mass Assembly from Large Scales to Small}
\vspace{-6pt}

Understanding how mass assembles in the Milky Way, from giant molecular clouds (GMCs) to dense, star-forming cores, is fundamental to understanding the star formation rate and constraining the physical origin of the stellar initial mass function (IMF), which in turn govern galaxy evolution through energy, momentum, and chemical feedback. GMCs, spanning hundreds of parsecs, serve as reservoirs of cold gas shaped by galactic environment, turbulence, and magnetic fields, and only a small fraction of this material ultimately forms stars \cite{Chevance2020}. Studies emphasize that local ambient galactic pressure coupled to feedback processes imprint directly on cloud properties, driving life-cycle dynamics at cloud scales \cite{Schruba2019}.

Sub-mm interferometers, though unable to probe large scale structure, have enabled systematic studies of the earliest stages of star cluster formation. The ALMA-IMF Large Program (LP) delivers a transformative dataset measuring the core mass function (CMF) across evolutionary stages, and preliminary results suggest little change in density concentration from young to evolved protoclusters, challenging models predicting strong dynamical evolution during early mass assembly \cite{Motte2022}. Evidence of inflowing streamers and detection of numerous hot cores indicates accretion and chemical complexity are already active across different scales \cite{Motte2022}. By providing a uniform, mass-complete census of cores in extreme environments, they establish a critical benchmark for testing theories of continual core growth versus fragmentation, the path to massive star formation \cite{Morii25}, and question the universality of the initial mass function (IMF) \cite{Pouteau22}. The ALMAGAL LP \cite[e.g.][]{Molinari25} leads the way in studying these high-mass star-forming cores across diverse Galactic environments.

While the IMF has been long treated as near-universal \cite{Bastian2010,Hopkins2018}, recent studies like these continue to question this premise, suggesting that environmental variation and sampling bias could obscure real differences in mass distributions \cite{Guszejnov2019}. These uncertainties further motivate detailed investigations into how inflow, fragmentation, magnetic regulation, and feedback vary with scale, environment, and time, all of which imprint upon the ultimate stellar census of star-forming regions.

The Milky Way offers the only laboratory where these processes can be spatially resolved across GMC, clump, filament, to star-forming core scales. Understanding mass assembly here is key to linking small-scale physics of star-formation to the equivalent properties of external galaxies, including star formation efficiency, cluster formation, and chemical and radiative enrichment of the ISM and circumgalactic medium.

With spectral lines, like those highlighted in Figure \ref{fig:overview}, we can study this mass assembly and structure formation across the many astrophysical scales relevant for Galaxy evolution: studying how clouds assemble mass, how that varies across physical size scales, and how the later stages of stellar lives feed back to the ambient clouds.  Different lines tracing different physical conditions tell us about the importance of mass assembly and feedback on all of these scales (from molecular clouds to cores to stars through structures such as clumps and filaments) with their velocity information breaking line of sight degeneracies and enabling the identification of coherent structures.

Getting into the details of quantifying mass assembly (via both spectral line observations and simultaneously observed continuum where possible) brings us to another strength of spectral line observations in the (sub-)mm: quantifying the physics and chemistry driving star formation through these physical scales.

\vspace{-12pt}
\subsection*{Physical and Chemical Conditions driving Star Formation}
\vspace{-6pt}

To constrain what regulates star formation on Galactic scales, and indeed, what shapes the IMF and local star forming efficiency requires an understanding of all of the phases of the ISM, the links between scales, and what physics dominates on each of those scales. The SKAO will, at longer wavelengths, be well placed to study the transition to molecular clouds. On the size scales of 10s - 0.1 pc and smaller is where (sub-)mm facilities are key for disentangling which physical (or chemical) processes dominate further mass assembly into colder, denser sub-structures.

Neutral atomic carbon ([C\,\textsc{i}]) is an excellent tracer of large-scale Galactic structure because it is wide-spread in molecular clouds, including regions where CO is weak or absent. [C\,\textsc{i}] is abundant in cloud envelopes and in CO-dark H$_2$ gas \cite{Beuther14}, making it sensitive to both dense and diffuse molecular material. Its relatively optically thin nature provides more accurate column density estimates and complements CO for mapping the carbon phase balance, making [C\,\textsc{i}] ideal for probing the distribution, evolution, and chemistry of molecular gas across the Galaxy.

On large scales, and at low temperatures and densities, molecules like CO are key tracers of gas motions, with higher energy transitions becoming more important as density or temperature increase. Other species such as CS, CN, HCN, N$_2$H$^+$ and HCO$^+$ remain optically thin at higher densities and are useful for tracing physical conditions at core boundaries, with the ionised species starting to give information on the ionisation state of the emitting medium and CN tracing the polarised emission linked to the magnetic fields. Higher excitation molecules like CH$_3$CN and other complex organic molecules (COMs) pinpoint where mass has assembled to the point of creating cores and thus where star formation is well underway (see also submissions by Sanchez-Monge et al., Rivilla et al., Jimenez-Serra et al. and K\"uffmeier et al.).

Some species are able to directly probe physical conditions. Sub-mm transitions of H$_2$CO are excellent densitometers (for n$\sim 10^4-10^7$ cm$^{-3}$, \cite{Ginsburg11}) and ratios of HCN/HNC are excellent thermometers (for T$\sim 10-50$K, \cite{Hacar20}). These densities and temperatures are typical of the dense molecular ISM and thus these simple probes can directly tell us a lot about the state of the gas. From those bedrocks, other species can be used to shape our more complex understanding of structure and flow in the dynamic ISM.

Submillimeter radio recombination lines (RRLs) are powerful probes of ionized gas in the Galaxy, offering insights into the structure and dynamics of H\,\textsc{ii} regions and the diffuse ionized medium. Their high-frequency nature reduces pressure broadening and opacity effects, providing accurate measurements of electron temperature, density, and kinematics. When combined with molecular tracers, sub-mm RRLs enable a multi-phase view of Galactic structure and star formation environments.

Observing these tracers of a wide range of physical and chemical parameters is key to building up a robust picture of mass assembly in our Galaxy. As shown in Figure 1, a number of studies have approached this problem over the past few decades, focusing on achievable area coverage and specific lines (like CO) because of the limits of current facilities. These heterogenous studies, while showing us what's possible, lack a consistency of sensitivity, spatial and spectral resolution and breadth of spectral probes.

Current and planned/funded facilities observing in the (sub-)mm regime suffer from a number of tradeoffs which make homogeneous spectral line mapping of the Galactic Plane impossible to any appreciable depth.  Current general purpose single-dish facilities are relatively limited by their available fields of view (FoV).  The LMT and IRAM 30m, two of the largest and most sensitive single dishes available, have FoV of order a few arcmin, which, even once fully populated by heterodyne receivers with many more elements than presently available, will severely limit the ability to undertake large molecular cloud surveying.  APEX and JCMT, while smaller, can operate at higher frequencies, however their focal planes are also still a limiting factor in survey speed.  At high resolution, ALMA, in its current form, also suffers from only being able to observe small fields of view and overall spatial filtering of large scale structure. 

Smaller, purpose built survey telescopes have the advantage of larger fields of view, but to achieve those fields, have had to compromise on resolution and sensitivity. Observatories like FYST will be able to observe high frequency sub-mm transitions of many lines, but at resolutions and sensitivities not well suited to studying all but the closest (i.e $<$0.5 kpc) star forming regions on the key size scales of clumps and cores.

To revolutionise our understanding of mass assembly in our Galaxy, and ensure that we are applying the right relations elsewhere requires a homogeneous study of a host of ionised and molecular line diagnostics that resolve all physical scales from GMCs (10s of pc) through filaments ($\sim$ a few pc), clumps ($\sim$ 0.3-3 pc) and cores ($\sim$ 0.1 pc). Observing across a full atmospheric transmission window in the (sub-)mm at spectral resolutions of order 0.1 km s$^{-1}$, reaching all of these physical size scales and doing so across the entire plane of the Galaxy visible from the Southern Hemisphere (the one best suited to Galactic Plane studies) requires new technologies. It needs a new single dish sub-mm facility capable of reaching resolutions of a few arcsec over large simultaneous fields of view in synergy with an enhanced ALMA to probe the size scales of star forming cores even out to the farthest (and lowest metallicity) reaches of the outer galaxy.

A full depth, high spectral resolution survey of the entire Galactic plane is not a good use of resource. Instead, we suggest a layered approach to balance depth and sky coverage and maximise scientific return. This exploits the complementary strengths of a single dish (AtLAST) and interferometer (ALMA) to combine a shallow, wide sweep of the Galactic Plane with deep integrations on targeted areas that bring the details into sharp focus while assuring that  the sample being drawn from is representative of the underlying population, not just the brightest objects in the sky.

A wide-area tier spanning $270^\circ$ of the Galactic plane ($b = \pm 1.5^\circ$) at $\sim 1.0\,\mathrm{K}$ sensitivity per $0.1\,\mathrm{km\,s^{-1}}$ channel, can provide a census of CO, [CI], and dense-gas tracers such as HCO$^+$ and CS, from which  CO spectral line energy densities (SLEDs), [CI]/CO ratios, and basic cloud properties can be derived. A medium tier narrows to $\pm 0.5^\circ$ and reaches $\sim 0.3\,\mathrm{K}$ sensitivity, enabling the densitometry and thermometry possible with H$_2$CO and HCN/HNC observations (respectively), and initial large scale measurements of complex organics. These two tiers would require a high-resolution single-dish telescope like AtLAST. Finally, a deep tier targets selected patches at $\sim 0.14\,\mathrm{K}$ to probe detailed chemistry and dynamics at higher resolution with overhauled ALMA. With all three tiers, we probe the physical properties of the gas, as well as its motions and capture, to various degrees, its chemical complexity. Through these studies of molecular clouds, we can trace the flow of gas through different physical scales, identify what enhances and inhibits that flow, and how the galactic lifecycle continues through returning enriched gas to the ISM.

\vspace{-12pt}
\section*{Technical Requirements}
\vspace{-6pt}

\textbf{AtLAST} is envisioned as a 50\,m dish on Chajnantor, delivering resolutions of a few arcsec across the submillimeter range (at 375 GHz, it reaches 4$''$, which at 4 kpc is $\sim$ 0.08 pc) and a field of view (FoV) of \textit{up to} 2$^\circ$ (at similar frequencies) for efficient Galactic plane sweeps and mapping of large star-forming regions. Equipped with large-format wideband heterodyne arrays of roughly $\mathcal{O}(10^3)$ pixels (which will not yet fill the FoV), it will offer 16--32\,GHz instantaneous bandwidth and 0.01--0.1\,km\,s$^{-1}$ spectral resolution, enabling simultaneous coverage of CO, [C\,\textsc{i}], dense-gas tracers, COMs and shock indicators. Its mapping speed will surpass current facilities by a few orders of magnitude, supported by scalable pipelines and archives designed for high (kilo-pixel~$\times$~GHz~$\times$~deg$^2$) data volumes. \textbf{ALMA2040}, as a complementary facility, will be transformative for the detailed tier of the survey assuming that the overall spectral response is increased by a factor of 10 (above the WSU) and it is upgraded to include sensitivity to larger scale structure. 

\vspace{6pt}

\textit{In summary, the combination of high throughput (large aperture and FoV) and high sensitivity of AtLAST is unprecedented for a facility that can access the whole (sub-)millimeter wavelength range. This is the only way we can undertake statistically robust studies of the composition of the gas in the Milky Way to know that the relationships we are building here are universal and/or applicable to other galaxies.  With spectral line surveys of the Galactic Plane, we unlock the full story of the ISM through chemistry and physics.}

\vspace{6pt}
\printbibliography[heading=none]

\end{document}

%% file: macros.tex
\definecolor{DarkGreen}{rgb}{0.0, 0.3, 0.0}
\definecolor{purple}{rgb}{0.5, 0.0, 0.5}
\definecolor{red}{rgb}{1, 0.0, 0.0}
\definecolor{green}{rgb}{0, 1.0, 0.0}

\def\3he{$^3{\rm He}$}

\def\lsim{\mathrel{\lower2.5pt\vbox{\lineskip=0pt\baselineskip=0pt
           \hbox{$<$}\hbox{$\sim$}}}}

\def\gsim{\mathrel{\lower2.5pt\vbox{\lineskip=0pt\baselineskip=0pt
           \hbox{$>$}\hbox{$\sim$}}}}